\documentstyle [11pt]{article}

\begin{document}
%
%
\newcommand{\Abs}[1]{|#1|}
\newcommand{\EqRef}[1]{(\ref{eqn:#1})}
\newcommand{\FigRef}[1]{fig.~\ref{fig:#1}}
\newcommand{\Abstract}[1]{\small
   \begin{quote}
      \noindent
      {\bf Abstract - }{#1}
   \end{quote}
    }
\newcommand{\FigCap}[2]{
\ \\
   \noindent
   Figure~#1:~#2
\\
   }
\newcommand{\beq}{\begin{equation}}
\newcommand{\eeq}{\end{equation}}
\newcommand{\csection}[1]{\begin{centering}

\vspace{0.5cm}

\section{#1}\end{centering}}
\newcommand{\csubsection}[1]
{\begin{centering}

\vspace{0.5cm}

\subsection{#1}\end{centering}}
\newcommand{\csubsubsection}[1]
{\begin{centering}

\vspace{0.5cm}

\subsubsection*{#1}\end{centering}}

%
%
%
\title{Decay of correlations, Lyapunov exponents and anomalous
diffusion in the
Sinai billiard\footnote{Talk to be presented at
{\em Nonlinear Dynamics, Chaotic and complex systems,
\mbox{NDCCS'95}}, Zakopane,
Poland, Nov. 1995}} %
\author{Per Dahlqvist \\
Mechanics Department \\
Royal Institute of Technology, S-100 44 Stockholm, Sweden\\[0.5cm]
}
\date{}
\maketitle
%
\Abstract{We compute the decay of the velocity
autocorrelation function, the Lyapunov exponent and the
diffusion constant for the
Sinai billiard within the framework of
dynamical zeta functions.
The asymptotic decay of the velocity autocorrelation
function is found to be $C(t) \sim c(R)/t$. The Lyapunov exponent for the
corresponding map agrees with
the conjectured limit
$\lambda_{map} \rightarrow  -2\log(R)+C$ as
$R \rightarrow 0$ where
$C=1-4\log 2+27/(2\pi^2)\cdot \zeta(3)$. The diffusion constant of the
associated Lorentz gas is found to be divergent
$D(t) \sim \log t$.
}


\csection{Theory}

\csubsection{Chaotic averages}

Consider a  chaotic Hamiltonian system with two degrees of freedom.
Assign a weight $w(x_0,t)$ to the trajectory starting at phase space point
$x_0$ and evolving during time $t$ (to point $x(x_0,t)$)
in such a way that it is multiplicative
along the flow: $w(x_0,t_1+t_2)=w(x_0,t_1)w(x(x_0,t_1),t_2)$.
The phase space average of $w(x_0,t)$ may be expanded in terms of periodic
orbits as \cite{PDlyap}
\begin{equation}
\lim_{t \rightarrow \infty}
\langle  w(x_0,t)\rangle  = \lim_{t \rightarrow \infty}
\sum_p T_p \sum_{r=1}^{\infty} w_p^r \frac{\delta(t-rT_p)}
{\Abs{(1-\Lambda_p^r)(1-1/\Lambda_p^r)}}
 \ \ ,
\label{eqn:tracedef}
\end{equation}
where $r$ is the number of repetitions of primitive orbit $p$, having period
$T_{p}$, and  $\Lambda_{p}$ is the expanding eigenvalue of the
Jacobian (transverse to the flow). $w_p$ is
the weight integrated along with cycle $p$.
Zeta functions are introduced by observing that the average
\EqRef{tracedef} may we written as
\begin{equation}
\lim_{t \rightarrow \infty}\langle w(x_0,t)\rangle  = \lim_{t \rightarrow
\infty} \frac{1}{2\pi i}
\int_{-i\infty+a}^{i\infty+a} e^{st}\frac{Z_w'(s)}{Z_w(s)}ds
\label{eqn:intlogder}   \ \ ,
\end{equation}
with the zeta function \cite{flow}

\begin{equation}
     Z_w(s)=\prod_{p}\prod_{m=0}^{\infty}
          \left(1-w_p \frac{e^{-sT_{p}}}
    {\Abs{\Lambda_{p}} \Lambda_{p}^{m}}\right)^{m+1}  \ \ .
\label{eqn:zetaw}
\end{equation}

The expressions are valid also for an area preserving map, like a Poincar\'{e}
map of the above system. The time vaiable $t$ and period $T_p$
are now discrete and
usually
denoted as $n$ and $n_p$. The zeta function \EqRef{zetaw} is usually expressed
in the $z=exp(s)$ plane. The Dirac delta in \EqRef{tracedef} should
be replaced by a Kronecker delta and the integral \EqRef{intlogder}
accordingly performed along
a circle $|z|=C$.

We will be interested in three weights of the required type.
In the discussion of {\em Lyapunov exponents}
we will be interested in
\begin{equation}
w_\lambda (x_0,t)=\Lambda(x_0,t )^\beta  \label{eqn:wlam}  \ \ ,
\end{equation}
where $\Lambda(x_0,t )$ is the (leading) eigenvalue of the Jacobian of the map
from $x_0$ to $x(x_0,t )$.
It is not exactly multiplicative but we will discuss that problem below.
We now express the Lyapunov exponent in terms of the associated
zeta function
\[
\lambda\equiv \lim_{t \rightarrow \infty} \frac{1}{t}
\langle \log |\Lambda(x_0,t)| \rangle
=\lim_{t \rightarrow \infty}\frac{1}{t}\frac{d}{d\beta}
\langle \Lambda(x_0,t)^\beta\rangle\mid_{\beta=0}
\]
\begin{equation}
=
\lim_{t \rightarrow \infty} \frac{1}{t}\frac{1}{2\pi i}
\int_{-i\pi+\epsilon}^{i\pi+\epsilon} e^{st} \frac{d}{d \beta}
\frac{d}{d s}
\log Z_\lambda(s) \mid_{\beta=0} ds \ \ .
\label{eqn:lamtrace}
\end{equation}

In the discussion of {\em diffusion} we will consider
\begin{equation}
w_{D}(x_0,t)=e^{\bar{\beta}\cdot(\bar{x}(x_0),t )-\bar{x}_0)}  \ \ ,
\label{eqn:wD}
\end{equation}
where $\bar{x}$ is the configuration space part of the phase space vector $x$.
The diffusion constant may now be expressed in terms of the associated zeta
function
\[
D=\lim_{t \rightarrow \infty}\frac{1}{2t}
\langle(\bar{x}(x_0),t )-\bar{x}_0)^2\rangle
=\lim_{t \rightarrow \infty}\frac{1}{2t}(\frac{d^2}{d\beta_1^2}
+\frac{d^2}{d\beta_2^2})\langle
e^{\bar{\beta}\cdot(\bar{x}(x_0),t )-\bar{x}_0)}\rangle\mid_{\beta=0} \]
\begin{equation}
= \lim_{t \rightarrow \infty}\frac{1}{2t} \frac{1}{2\pi i}
\int_{-i\infty+\epsilon}^{i\infty+\epsilon} e^{st}
(\frac{d^2}{d\beta_1^2}+\frac{d^2}{d\beta_2^2})
\frac{d}{d s}
\log Z_D(s) \mid_{\bar{\beta}=\bar{0}} ds   \ \ .
\label{eqn:Dtrace}
\end{equation}

Next assume that we have a Poincar\'{e} map defined by a surface of section
$\Omega_{s.o.s.}$.
Then, using weight
\begin{equation}
e^{-s\cdot T(x_0,n)}w(x_0,n) \ \ , \label{eqn:w3}
\end{equation}
where $T(x_0,n)$ and $w(x_0,n)$ is the integrated time and weight
along a trajectory starting at $x_0 \in\Omega_{s.o.s.}$ and evolving during
$n$ iterates of the map.
We thus get a zeta function which combines the description of the map and the
flow.
\begin{equation}
     Z(s,z)=\prod_{p}\prod_{m=0}^{\infty}
          \left(1-w_p z^{n_p}e^{-sT_{p}}\frac{1}
    {\Abs{\Lambda_p} \Lambda_{p}^m}\right)^{m+1}  \ \ .
 \label{eqn:Zsz}
\end{equation}
A more lucid introduction to chaotic averages may be found in
\cite{PCLA} with proper references.

\csubsection{Approximate zeta functions}

Given a time continous system, suppose we have defined a surface
of section in such a way that the corresponding map destroys all
information from one iterate to another. Or more precisely, the weight
associated with consecutive iterates are uncorrelated:
$\langle w_i\;w_{i+1}\rangle=\langle w_i\rangle\langle w_{i+1}\rangle$.
This is called {\em Assumption A} in ref. \cite{BER}.
This can of course not be exactly fulfilled in practice. But suppose
our system is intermittent and we put the s.o.s. in the border between
the laminar and the chaotic phase. Then the chaotic phase will do its best
to destroy the memory from one laminar phase to the next.
We will now average the the weight over the surface of section.
\begin{equation}
\langle e^{-sT(x_0,n)}w(x_0,n)\rangle_{s.o.s.}=
\langle e^{-sT(x_0,1)}w(x_0,1)\rangle^n_{s.o.s.} \ \ ,
\end{equation}
which follows from assumption A.
This type of behaviour  is equivalent to their being only one zero
$z_0(s)=\langle \exp(-sT(x_0,1))w(x_0,1)\rangle_{s.o.s.}$
of $Z(s,z)$ in the $z-$ plane.
In the language of Ruelle resonances this means that there is an infinite
gap to the first resonance.
This in turn implies that the $Z(s,z)$ may be written.
\begin{equation}
Z(s,z)=e^{h(z,s)}\left(z-\langle e^{-sT(x_0,1)}w(x_0,1)
\rangle_{s.o.s.} \right)  \ \ ,
\end{equation}
where $h(z,s)$ is entire in $z$ and $s$.
The zeta function for the flow is obtained as
$Z_{flow}(s)=Z(z=1,s)$, see eq. \EqRef{Zsz}, and we get
\begin{equation}
Z_{flow}(s)=1-\langle e^{-sT(x_0,1)}w(x_0,1)\rangle_{s.o.s.}  \ \ .
\label{eqn:Zflow} \end{equation}
We had to put $h=(z=1,s)=0$ in order for Z(s) to obtain the correct limit at
infinity. 
Normally, the best one can hope for is an finite gap
to the leading resonance
of the Poincar\'{e} map. The zeta function above is then only approximate,
indicated by putting a hat on it:
$\hat{Z}$,
we also
suppress the clumsy subscript {\em flow}.
The first two terms in an expansion around $s=0$
for the special case $w=1$
are however exact:
$\hat{Z}(s=0)=Z(s=0)=0$ and $\hat{Z}'(s=0)=Z'(s=0)$ for linearity reasons.

We can reformulate the formula above for the special case where $w(x_0,1)$ is a
function of $T(x_0,1)\equiv\Delta(x_0)$ denoted $w(\Delta)$
Then the probability distribution of recurrence times is
\begin{equation}
p(\Delta)=\langle \delta(\Delta-\Delta(x_0))\rangle_{s.o.s.} \ \ ,
\end{equation}
and
\begin{equation}
\hat{Z}(s)=1-\int_0^{\infty}w(\Delta)p(\Delta)e^{-s\Delta}d\Delta  \ \ .
\end{equation}

Approximate zeta functions were introduced in refs. \cite{PDreson,PDsin},
to a large extent inspired by ref. \cite{BER}, the above derivation is
though much more easy to follow than the one in \cite{PDreson}.

\csection{The Sinai Billiard}

We will apply the method of approximate zeta functions to
the Sinai billiard\cite{Sin},
a particle with unit speed bouncing inside a unit
square with a scattering circle of radius $R$ in the middle.
The only sensible choice of $\Omega_{s.o.s.}$ is the disk itself.
The disk to disk map will be hyperbolic but without
a finite Markov partition, most likely it will have a finite gap and
exhibit exponential mixing \cite{Liv,RA}.
In the small scatterer limit one may derive the following exact expression for
the distribution of recurrence times \cite{PDsmall}.

\[
p_{R\rightarrow 0}(\Delta)
\]
\begin{equation}
= \left\{ \begin{array}{ll}
\frac{12R}{\pi^2} & \xi<1\\
\frac{6R}{\pi^2 \xi^2}(2\xi+\xi(4-3\xi)\log(\xi)+
4(\xi-1)^2 \log(\xi-1)-(2-\xi)^2\log|2-\xi|)&
\xi>1 \end{array} \label{eqn:psmall} \right.  \ \ ,
\end{equation}
where
$\xi=\Delta 2R$
In the following we will e.g. need
the mean $<\Delta>=1/2R$ in this distribution, and the large $\Delta$ limit:
$p \sim 1/(\pi^2R^2\Delta^3)$.

\csubsection{The Lyapunov exponent}

Consider a trajectory with $n$ intersections with $\Omega_{s.o.s.}$.
The consecutive disk to disk time of flight and scattering angles will
be denoted as $\Delta_i$ and $\alpha_i$ being functions of the s.o.s coordinate
$x_i$.
It may be shown \cite{PDsmall} that
the (by construction)
multiplicative weight
\begin{equation}
w(x_0,n)=\prod_{i=1}^{n} \frac{2\Delta_i}{R\cos \alpha_i}   \ \ ,
\label{eqn:wfudge}
\end{equation}
give the correct $R \rightarrow 0$ limit of the Lyapunov exponent
\begin{equation}
\lambda_{map}=\lim_{n\rightarrow \infty}\frac{1}{n}
\langle \log w(x_0,n) \rangle_{s.o.s}+O(R)=
\langle \log \frac{2\Delta(x)}{R\cos \alpha(x)}
\rangle_{s.o.s}+O(R) \end{equation}
\[
=\log(2/R)+\langle \log \Delta(x)\rangle_{s.o.s}-
\langle \log \cos \alpha(x) \rangle_{s.o.s}+O(R)  \ \ .
\]
The problem is then reduced to computing
the average of $\log \Delta$  (the average of $\cos \alpha$ is trivial).
To that end we use eq. \EqRef{psmall}.
The result is \cite{PDsmall}
\begin{equation}
\lambda_{map} \rightarrow -2\log R+C  \;  \;  \;  \;
R\rightarrow \infty  \ \ ,
\end{equation}
with $C=1-4\log 2+27/(2\pi^2)\cdot \zeta(3)$, confirming previous
conjectures \cite{Oono,Bouch,Chern} but with $C$ computed exactly for the first
time.

What about the Lyapunov exponent of the flow.
The zeta function is now, cf. eqs. \EqRef{wlam} and \EqRef{Zflow},
\begin{equation}
\hat{Z}=1-\langle \Lambda(x_o)^\beta e^{-s\Delta(x_0)} \rangle_{s.o.s}=
s\langle \Delta \rangle +O(s^2\log s)-\beta
\left(\lambda_{map}+O(s) \right) \ldots
\end{equation}
Putting this into eq. \EqRef{lamtrace} we get
\[
\lambda=
\lim_{t \rightarrow \infty} \frac{1}{t}\frac{1}{2\pi i}
\int_{-i\infty+\epsilon}^{i\infty+\epsilon} e^{st}
(\frac{\lambda_{map}}{\langle \Delta \rangle s^2}) (1+O(s\log s)) ds
\]
\begin{equation}
=\lim_{t \rightarrow \infty} \frac{1}{t}
(\frac{\lambda_{map}}{\langle \Delta \rangle}t+O(\log t))=
\frac{\lambda_{map}}{\langle \Delta \rangle}  \ \ ,
\end{equation}
which is just Abramovs formula \cite{Abr}.
We got an exact result (for the weight \EqRef{wfudge})
since the relevant terms in the approximate zeta function were indeed
exact.
We saw that the leading zero also being a branch point did not do any harm
to the Lyapunov exponent but it will in fact induce a phase transition among
the generalized Lyapunov exponents \cite{PDlyap}.

\csubsection{Diffusion}

Next we consider the associated Lorentz gas obtained by infinitely
unfolding the billiard. The zeta function is, cf. eqs. \EqRef{wD} and
\EqRef{Zflow}
\begin{equation}
\hat{Z}(s)=1-\langle e^{-s\Delta}e^{x_1\beta_1+x_2\beta_2} \rangle=
1-\langle e^{-s\Delta}\frac{1}{2\pi}\int_0^{2\pi}
e^{\Delta\beta cos(\theta)}d\theta  \rangle  \ \ ,
\end{equation}
where we used the asymptotic isotropy (which is not really necessary cf
refs. \cite{PDlyap,Bleh}), and $\Delta=\sqrt{x_1^2+x_2^2}$.
We only need to expand to
second order in $\beta$
\[
\hat{Z}(s)=1-\langle e^{-s\Delta} \rangle -\frac{\beta^2}{4}
\langle \Delta^2 e^{-s\Delta} \rangle \ldots
\]
\begin{equation}
=1-\frac{s}{2R} +O(s^2 \log s) +\frac{\beta^2}{4\pi^2R^2}
(-\log s+O(1)) \ldots   \label{eqn:ZDexpand}  \ \ .
\end{equation}
Putting it into eq \EqRef{Dtrace} we get
\begin{equation}
D =\lim_{t \rightarrow} \frac{1}{2t} \frac{2}{\pi^2R}
\frac{1}{2\pi i}
\int_{-i\infty+\epsilon}^{i\infty+\epsilon} e^{st}\frac{-\log s +O(1)}{s^2} ds
=\frac{1}{\pi^2R}(\log(t)+O(1))  \ \ ,
\end{equation}
which is indeed the exact result \cite{Bleh}.
Exact result may also be obtained for finite $R$ \cite{PDlyap,Bleh}.
The exactness of the result indicates that the term
$-\frac{\beta^2}{4\pi^2R^2}\log s$ in eq. \EqRef{ZDexpand}
agrees with the corresponding one in the exact zeta function.

\csubsection{Correlation functions}

As regarding correlation functions we will restrict ourselves
to observables $A(x_0)$
that are changed only by bounces on the disk, such as
$A=|v_x|$.
The autocorrelation function is e.g. obtained as the time average
\begin{equation}
C(t)=\langle A(t_0+t)A(t_0) \rangle_{t_0}-\langle A^2 \rangle  \ \ .
\end{equation}
Measuring $A$ at two points of time separated by $t$, then
there is a probability $p_0(t)$ that the trajectory has {\em not} hit
the disk in between and according to assumption A we get \cite{DAcorr}
\begin{equation}
C(t)=p_0(t)\langle A^2 \rangle+(1-p_0(t))\langle A \rangle^2
-\langle A^2 \rangle=
p_0(t)V(A)   \ \ , \label{eqn:nice}
\end{equation}
where $V(A)$ is the variance of $A$.
The function $p_0(t)$ may be expressed in terms of $p(\Delta)$ \cite{DAcorr}
\begin{equation}
p_0(t)=\frac{1}{\langle \Delta \rangle}
\int_t^\infty \{ \int_u^\infty p(\Delta)d\Delta \} du  \ \ .
\label{eqn:p0}
\end{equation}
The $1/\Delta^3$ decay of $p(\Delta)$ thus implies a $1/t$ of the
correlation function.
In fig 1 we compare the numerical correlation function with $A=|v_x|$
with eq \EqRef{nice} with $p_0(t)$ computed from eq. \EqRef{psmall}.
The $1/t$ decay law has also been suggested in refs \cite{Frid,RA}

I would like to acknowledge Roberto Artuso with whom I cooperate
on the correlation decay problem.
This work was supported by the Swedish Natural Science
Research Council (NFR) under contract no. F-FU 06420-303.

\vspace{0.5cm}

\newpage

\newcommand{\PR}[1]{{Phys.\ Rep.}\/ {\bf #1}}
\newcommand{\PRL}[1]{{Phys.\ Rev.\ Lett.}\/ {\bf #1}}
\newcommand{\PRA}[1]{{Phys.\ Rev.\ A}\/ {\bf #1}}
\newcommand{\PRD}[1]{{Phys.\ Rev.\ D}\/ {\bf #1}}
\newcommand{\PRE}[1]{{Phys.\ Rev.\ E}\/ {\bf #1}}
\newcommand{\JPA}[1]{{J.\ Phys.\ A}\/ {\bf #1}}
\newcommand{\JPB}[1]{{J.\ Phys.\ B}\/ {\bf #1}}
\newcommand{\JCP}[1]{{J.\ Chem.\ Phys.}\/ {\bf #1}}
\newcommand{\JPC}[1]{{J.\ Phys.\ Chem.}\/ {\bf #1}}
\newcommand{\JMP}[1]{{J.\ Math.\ Phys.}\/ {\bf #1}}
\newcommand{\JSP}[1]{{J.\ Stat.\ Phys.}\/ {\bf #1}}
\newcommand{\AP}[1]{{Ann.\ Phys.}\/ {\bf #1}}
\newcommand{\PLB}[1]{{Phys.\ Lett.\ B}\/ {\bf #1}}
\newcommand{\PLA}[1]{{Phys.\ Lett.\ A}\/ {\bf #1}}
\newcommand{\PD}[1]{{Physica D}\/ {\bf #1}}
\newcommand{\NPB}[1]{{Nucl.\ Phys.\ B}\/ {\bf #1}}
\newcommand{\INCB}[1]{{Il Nuov.\ Cim.\ B}\/ {\bf #1}}
\newcommand{\JETP}[1]{{Sov.\ Phys.\ JETP}\/ {\bf #1}}
\newcommand{\JETPL}[1]{{JETP Lett.\ }\/ {\bf #1}}
\newcommand{\RMS}[1]{{Russ.\ Math.\ Surv.}\/ {\bf #1}}
\newcommand{\USSR}[1]{{Math.\ USSR.\ Sb.}\/ {\bf #1}}
\newcommand{\PST}[1]{{Phys.\ Scripta T}\/ {\bf #1}}
\newcommand{\CM}[1]{{Cont.\ Math.}\/ {\bf #1}}
\newcommand{\JMPA}[1]{{J.\ Math.\ Pure Appl.}\/ {\bf #1}}
\newcommand{\CMP}[1]{{Comm.\ Math.\ Phys.}\/ {\bf #1}}
\newcommand{\PRS}[1]{{Proc.\ R.\ Soc. Lond.\ A}\/ {\bf #1}}
%


\newpage

\section*{Figure captions}

\FigCap{1}{Numerical correlation function (full line), for $R=0.106$.
The dashed
line  represents eq. \EqRef{nice} with $p_0(t)$ computed from eqs.
\EqRef{p0} and \EqRef{psmall}. From ref. \cite{DAcorr}}


\begin{thebibliography}{99}
%
{\small
\bibitem{PDlyap} P.~Dahlqvist, {\em Lyapunov exponents and anomalous
                 diffusion of a Lorentz gas with
                 infinite horizon using approximate zeta functions},
                 submitted to \JSP{}.
\bibitem{flow} P.~Cvitanovi\'{c} and B.~Eckhardt,
                {\em Periodic orbit expansions for classical smooth flows},
                \JPA{24}, L237, (1991).
\bibitem{PCLA}  P. Cvitanovi\'c, {\em Dynamical averaging in terms of
                periodic orbits}, Physica {\bf D83}, 109
                (1995).
\bibitem{PDreson}P.~Dahlqvist, {\em Determination of resonance spectra
                 for bound chaotic systems}, \JPA{27}, 763 (1994).
\bibitem{PDsin} P.~Dahlqvist, {\em Approximate zeta functions for the
                Sinai billiard and related systems}, Nonlinearity {\bf 8},11
(1995).
\bibitem{BER}   V.~Baladi, J.~P.~Eckmann and D.~Ruelle,
                {\em Resonances for intermittent systems},
                Nonlinearity {\bf 2}, 119, (1989).
\bibitem{Sin}   Y.~G.~Sinai, {\em Dynamical Systems with Elastic reflections},
                Russ.\ Math.\ Surv. {\bf 25}, 137, (1979).
\bibitem{Liv}    C.~Liverani, {\em Decay of correlations}, Roma II preprint.
\bibitem{RA}     R.~Artuso, Giulio.~Casati and I.~Guarneri, {\em Numerical
                 experiments on Billiards}, Como preprint (1995).
\bibitem{PDsmall}P.~Dahlqvist, {\em The Lyapunov exponent in the
                Sinai billiard in the small scatterer limit}, in preparation.
\bibitem{Oono}  B.~Friedman, Y.~Oono and I.~Kubo, {\em Universal behaviour
                of Sinai billiard systems in the small-scatterer limit},
                \PRL{52}, 709, (1984).
\bibitem{Bouch} J-P.~Bouchaud and P.~L.~Doussal, {\em Numerical study
                 of a D-dimensional periodic Lorentz gas with
                 universal properties}, \JSP{41}, 225, (1985).
\bibitem{Chern}  N.~I.~Chernov, Funct.\ Anal.\ Appl.\ {\bf 25}, 204, (1991).
\bibitem{Abr}   L.~M.~Abramov, Dokl.\ Akad.\ Nauk.\ SSSR {\bf 226},
                128, (1959).
\bibitem{Bleh}  P.~M.~Bleher, {\em Statistical Properties of
                Two-Dimensional Periodic Lorentz Gas with Infinite Horizon}
                \JSP{66}, 315, (1992).
\bibitem{DAcorr} P.~Dahlqvist, R.~Artuso, {\em Decay of correlations in the
Sinai
                 billiard}, in preparation.
\bibitem{Frid}  B.~Friedman and R.~F.~Martin, \PLA{105}, 2 (1984).

}
\end{thebibliography}
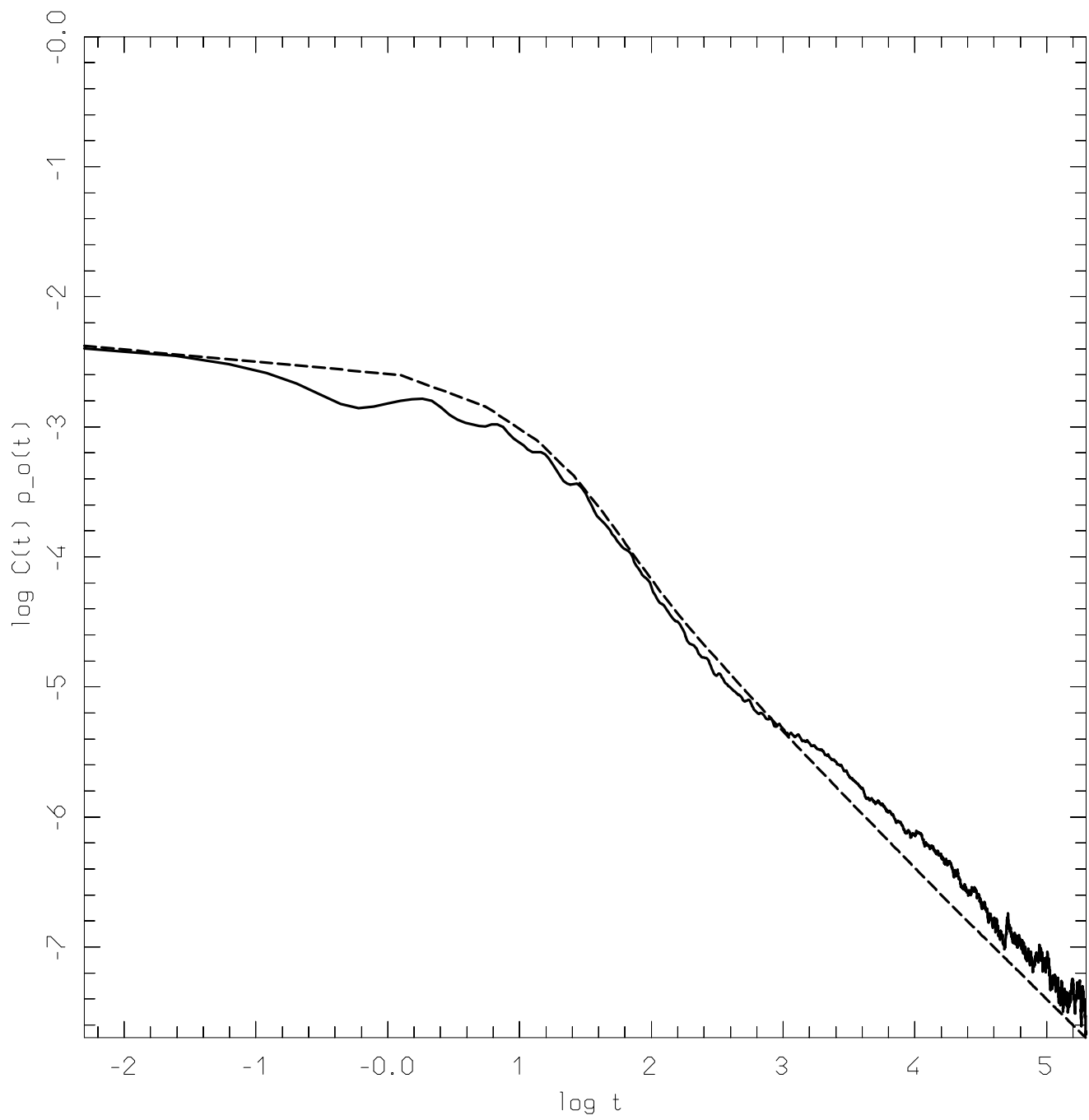
\end{document}